\documentclass[a4paper,11pt]{article}
\usepackage{pos}

\title{Particles in a pocket}

\author*[a]{Kirill Skovpen}

\affiliation[a]{Ghent University,\\
  Proeftuinstraat 86, B-9000 Gent, Belgium}

\emailAdd{kirill.skovpen@ugent.be}

\abstract{Communicating science through mobile smartphone and tablet
applications is one of the most efficient ways to reach general public
of diverse background and age coverage. The Higgsy project was created
in 2022 to celebrate the 10th anniversary of the discovery of the Higgs boson
at CERN. This project introduces a mobile game to search for
the Higgs boson production in a generic particle detector. The
MatterBricks is an augmented-reality project that was created for a major national event in
Belgium, held in 2023. The main features of the two mobile applications 
and further prospects for reaching general public through mobile application
development process are discussed.}

\FullConference{The European Physical Society Conference on High Energy Physics (EPS-HEP2023)\\
 21-25 August 2023\\
Hamburg, Germany\\}

\begin{document}
\maketitle

\section{Introduction}

If you are not reading a paper book while taking public transport, you are
probably staring at your mobile phone, or sleeping. Mobile devices are everywhere
these days providing us with communication means with our friends and
relatives, daily news, blogs, entertainment, and much more. These small ingenious
inventions can be easily put into one's pocket to carry along the
superpowers and wisdom of past generations. Topics related to fundamental scientific
research are not among the most popular things our society regularly
researches on the net. While this observation can be simply an intrinsic property
of the society, we think that it worth the candle to show it once
again that the fundamental research connects to truly fascinating
things that can not be overlooked.

\begin{figure}[hbtp]
\begin{center}
\includegraphics[width=0.23\linewidth]{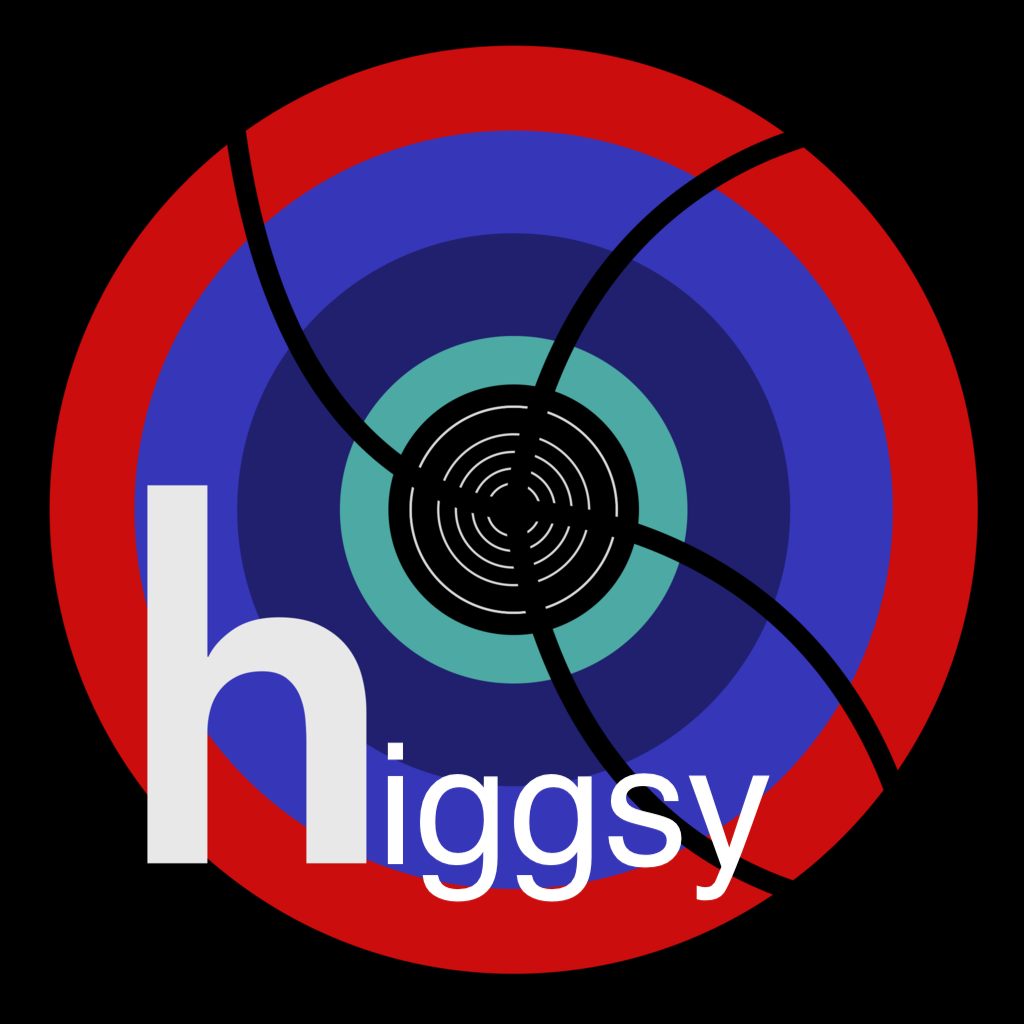}
\includegraphics[width=0.23\linewidth]{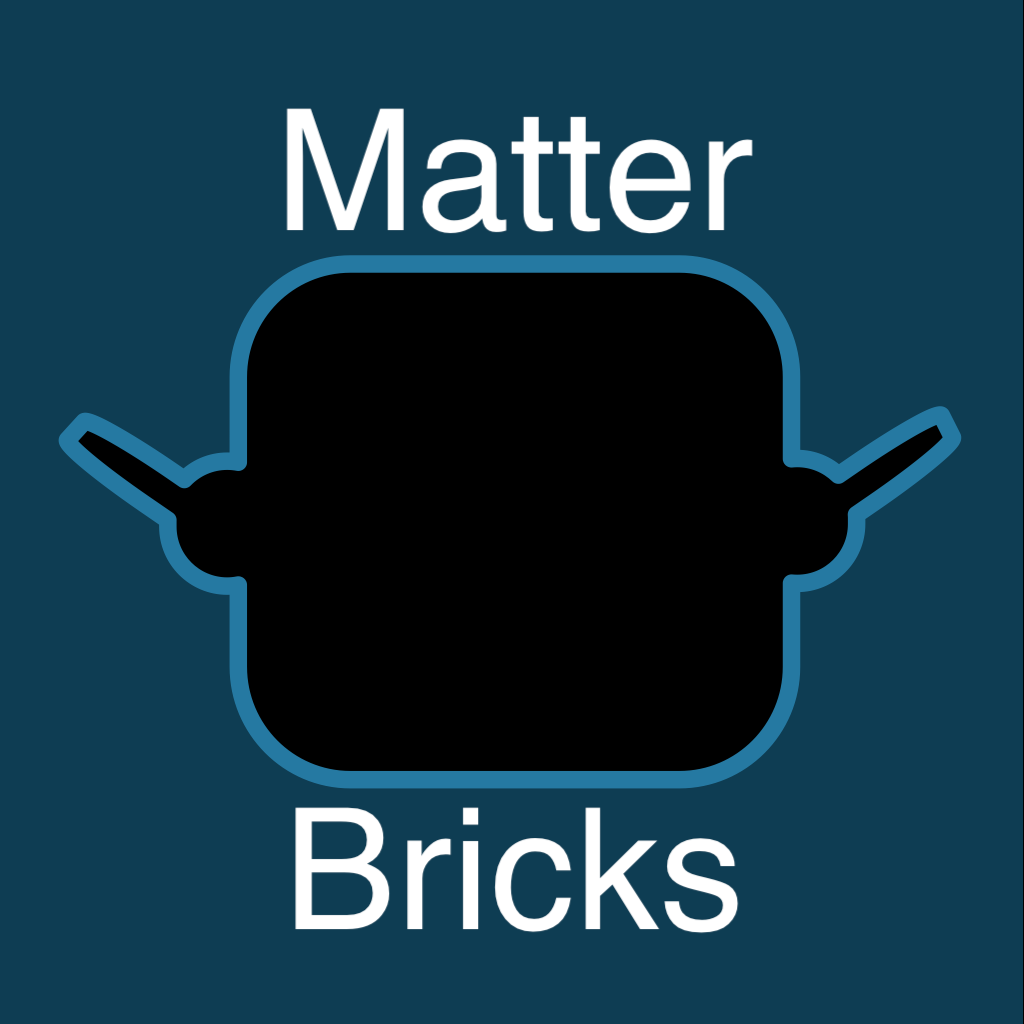}
\caption{The logo images of the Higgsy (left)~\cite{Higgsyg} and
MatterBricks (right)~\cite{MatterBricksg} projects.}
\label{fig:logo}
\end{center}
\end{figure}

Creations that are driven by scientific advancements in the field
of particle physics and related research areas include organization of
masterclasses~\cite{Masterclass,Masterclass2,Masterclass3}, gaming
experiences~\cite{GWgame,Qgame,QCDgame,PCgame},
demonstrator projects~\cite{CosmicWatch,CRdet}, professional
applications~\cite{PDGapp,EDatlas}, etc. Many outreach studies are performed within the International Particle
Physics Outreach Group (IPPOG)~\cite{IPPOG}. Development of science-popularizing applications that can
be installed on a handheld device is a very efficient method to reach diverse populations from
different cultural, racial, educational, and social backgrounds,
anywhere, anytime. In this work, we present Higgsy and
MatterBricks mobile games (Fig.~\ref{fig:logo}) that were developed for the iOS operating
system~\cite{iOS}, inspired by the rich world of particle physics.

\section{Higgsy}

\begin{figure}[hbtp]
\begin{center}
\includegraphics[width=0.60\linewidth]{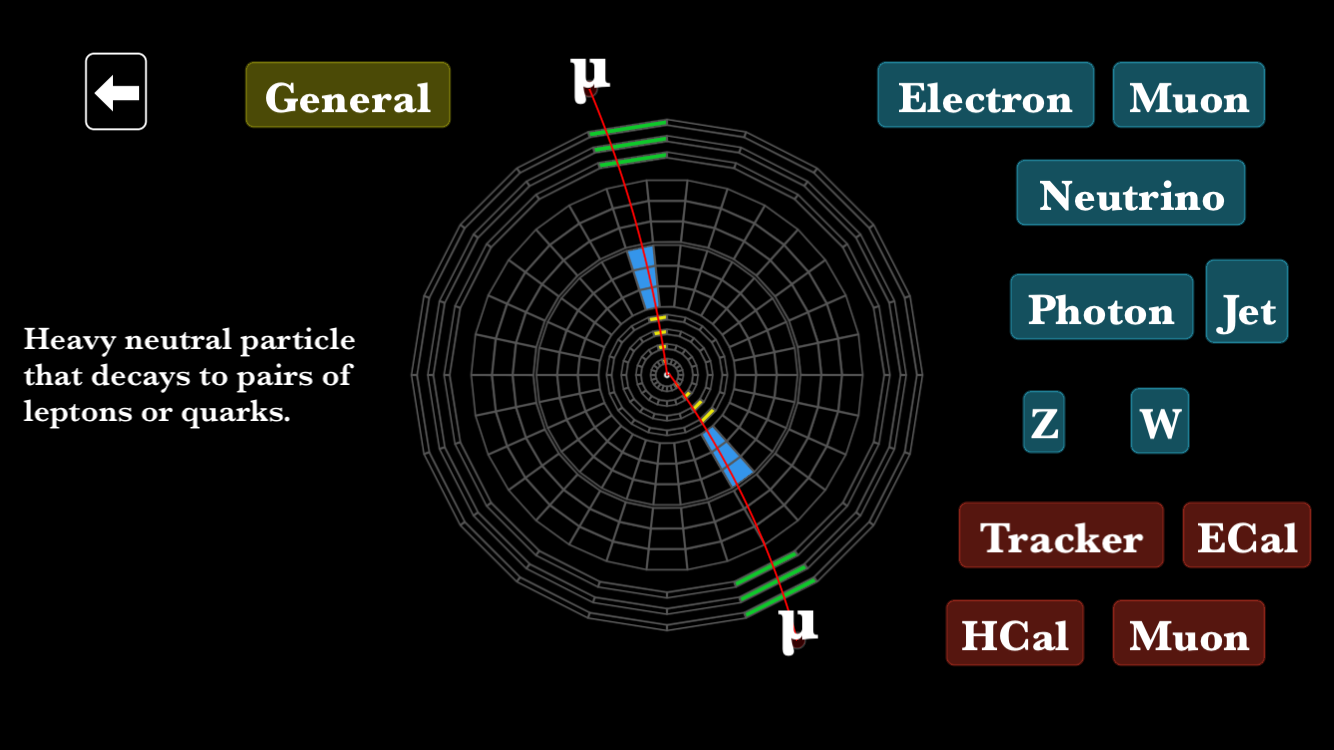}
\includegraphics[width=0.60\linewidth]{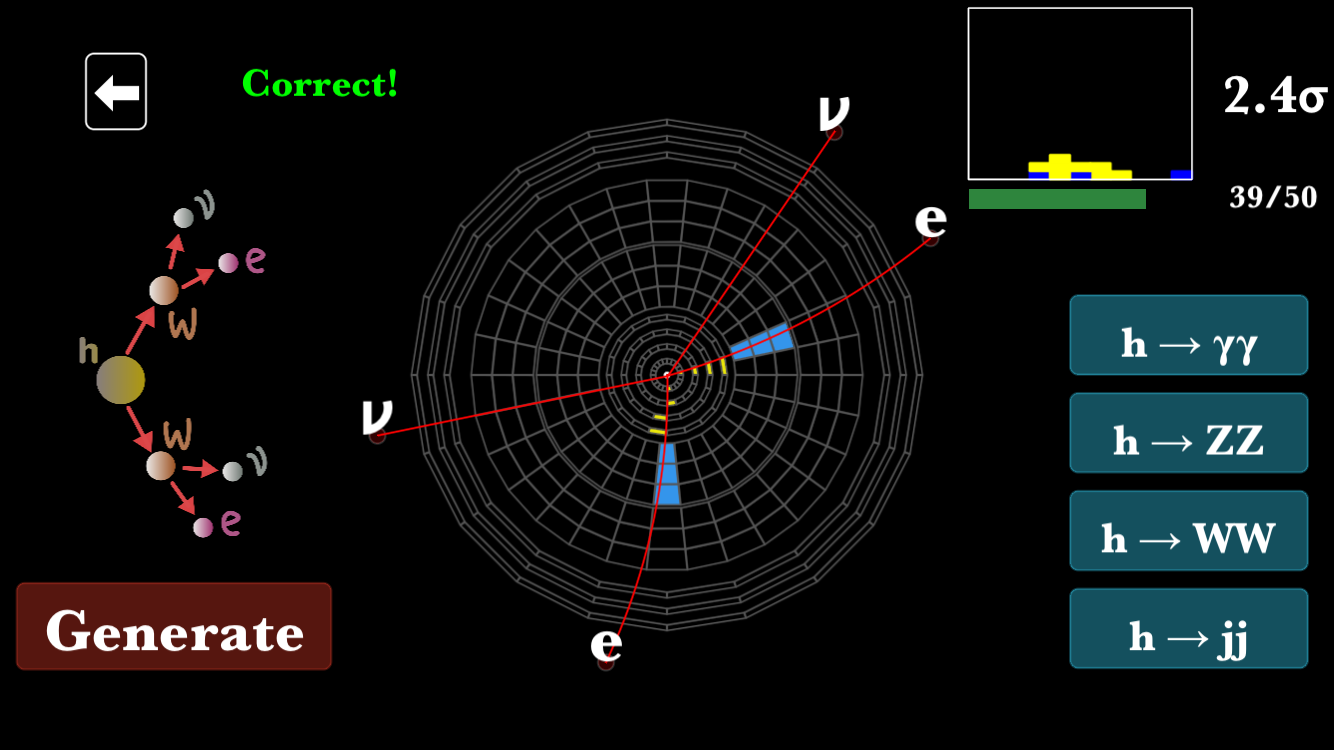}
\caption{Screen captures of Higgsy showing learning (left)
and Higgs-hunting (right) modes of the gameplay.}
\label{fig:higgsy}
\end{center}
\end{figure}

The discovery of the Higgs boson at the Large Hadron Collider (LHC) at
CERN marked a scientific breakthrough in our understanding of fundamental
interactions included in the standard model (SM) of particle
physics~\cite{HiggsATLAS,HiggsCMS}. This outstanding scientific
achievement was celebrated at its 10th anniversary in 2022 at
CERN~\cite{HiggsATLAS10,HiggsCMS10}. The Higgsy project was created to
relive the unforgettable experience of discovering the Higgs boson at
the LHC and make it accessible to everyone~\cite{Higgsyg}. The gameplay
includes several interactive gaming modes, explaining the main
features of the proton-proton collisions at the LHC,
and inviting a player to participate in an actual hunt for the Higgs boson. 
The player can generate elementary 
particles and their decays to study the detector-level information
arising from the interactions of these particles with the material of
the detector. This learning gaming phase allows the player to become familiar with
elementary particles and associate them with different types of
interactions, appreciating experimental challenges in properly
identifying a certain type of events. Once familiar with the
contents of the game, the player can make an attempt to properly
identify a required number of events with the Higgs boson production
in order to reach a statistically significant observation. The
learning and Higgs-hunting phases of the game are illustrated in
Fig.~\ref{fig:higgsy}.

\section{MatterBricks}

\begin{figure}[hbtp]
\begin{center}
\includegraphics[width=0.40\linewidth]{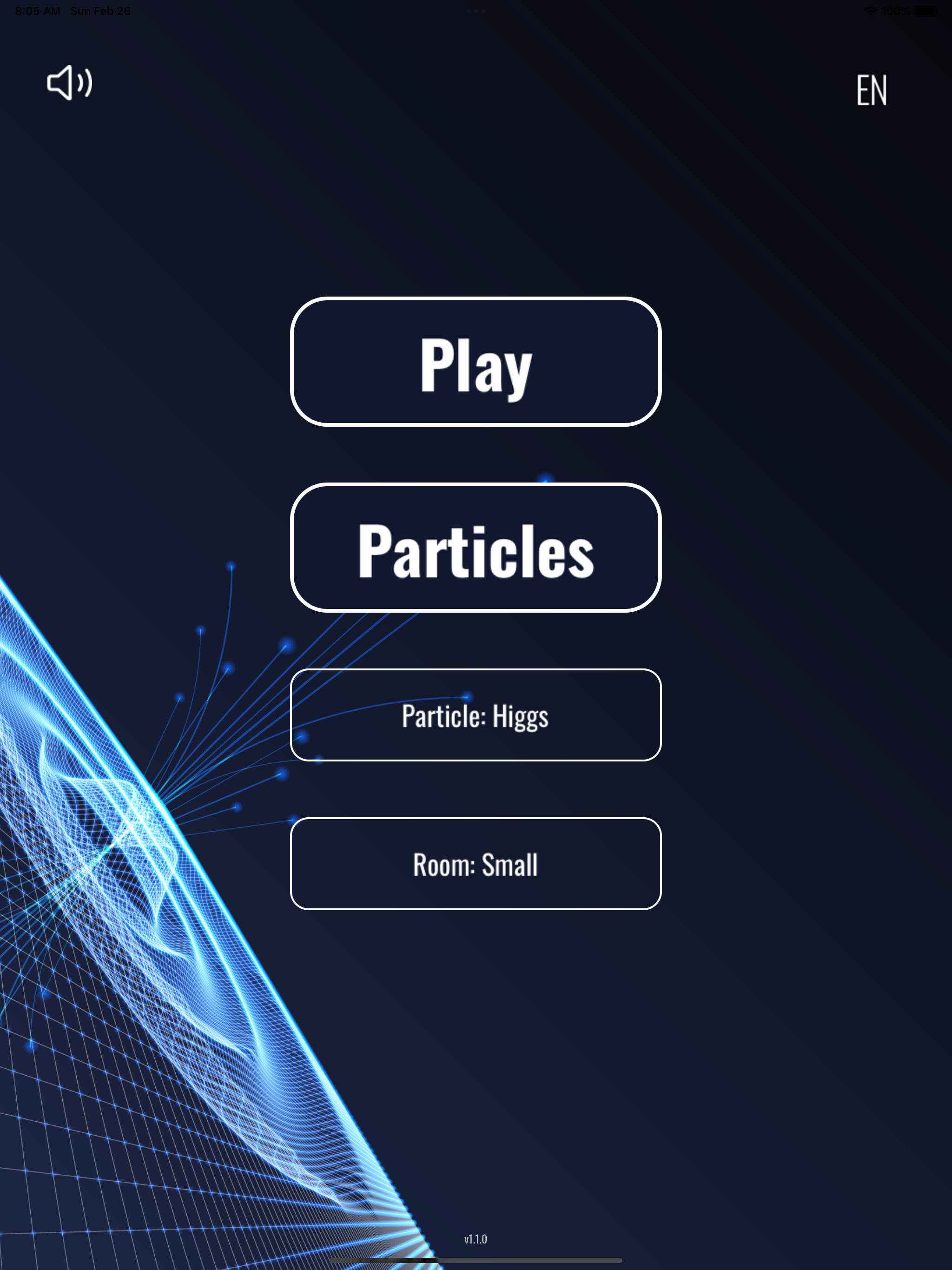}
\includegraphics[width=0.40\linewidth]{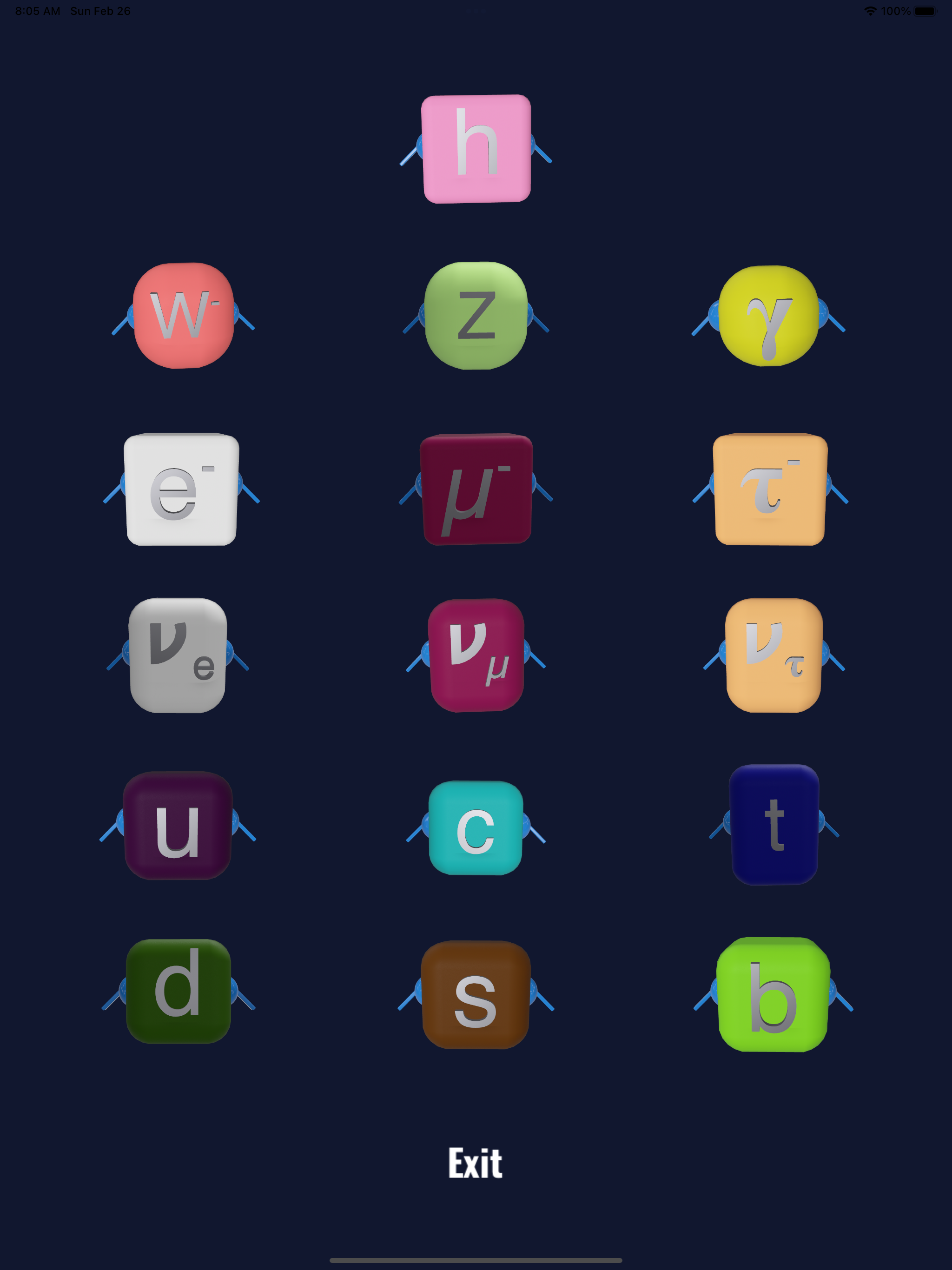}
\caption{Screen captures of the main menu of MatterBricks.}
\label{fig:matterbricks1}
\end{center}
\end{figure}

\begin{figure}[hbtp]
\begin{center}
\includegraphics[width=0.30\linewidth]{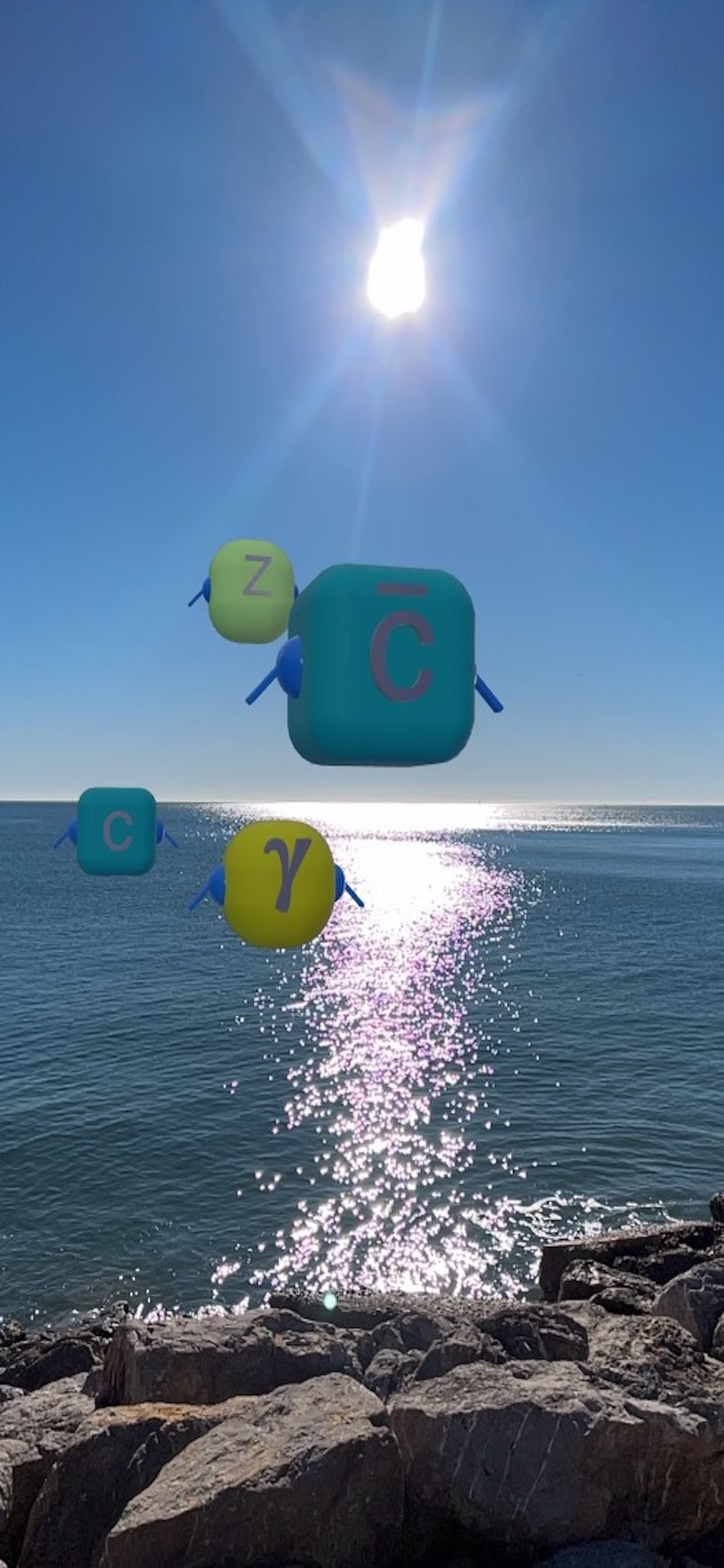}
\includegraphics[width=0.30\linewidth]{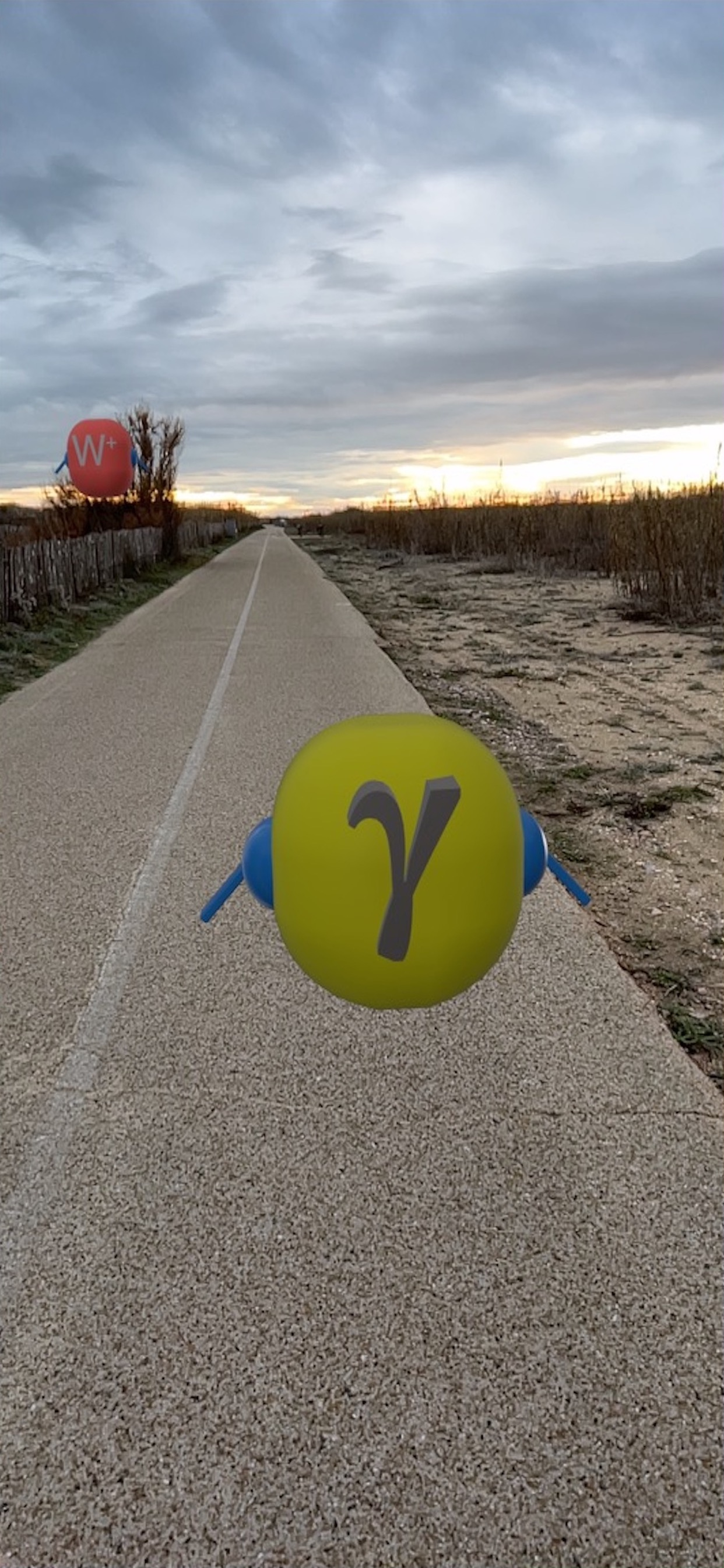}
\caption{Virtual particles projected onto real surroundings with the
help of augmented reality.}
\label{fig:matterbricks2}
\end{center}
\end{figure}

Novel technologies using virtual and augmented reality (AR) digital experiences have
been extremely successful in significantly extending our real-world
environment to unexplored territories. We decided to populate these
unknown worlds with elementary particles created with
MatterBricks~\cite{MatterBricksg} for the open symposium in
Belgium~\cite{DoD}. The player gets introductory 
explanations about these particles (Fig.~\ref{fig:matterbricks1}) to then dive into the world
augmented with the products of their decays
(Fig.~\ref{fig:matterbricks2}). The goal of the game is
to reassemble pairs of particles into their initial particle-origin.
As some say, ``gotta catch'em all''.

\section{Summary and outlook}

It's better to see something once, than to hear about it a thousand
times. If you haven't installed Higgsy and MatterBricks on your phone or
tablet, you should do it now. These small applications will fill you
with joy and hunger for more science. If they really do, we have accomplished
our mission. Both Higgsy and MatterBricks can be introduced in a
classroom, played outside, or simply discovering it on their own.
Our future work includes the development of similar applications for other mobile platforms, 
such as Android~\cite{Android}, in order to reach a broader audience.

\end{document}